\RequirePackage{lineno}
\documentclass[%
reprint,
superscriptaddress,
showpacs,preprintnumbers,
 amsmath,amssymb,
 aps,
longbibliography,
floatfix,
]{revtex4-1}

\usepackage{graphicx}
\usepackage{dcolumn}
\usepackage{bm}
\usepackage{xspace}	
\usepackage{color}
\usepackage{xcolor}
\usepackage{float}
\usepackage{amsmath}
\usepackage{verbatim}
\usepackage{appendix}
\usepackage{multirow}
\usepackage[colorinlistoftodos]{todonotes}

\newcommand*\patchAmsMathEnvironmentForLineno[1]{%
   \expandafter\let\csname old#1\expandafter\endcsname\csname #1\endcsname
   \expandafter\let\csname oldend#1\expandafter\endcsname\csname end#1\endcsname
   \renewenvironment{#1}%
      {\linenomath\csname old#1\endcsname}%
      {\csname oldend#1\endcsname\endlinenomath}}%
\newcommand*\patchBothAmsMathEnvironmentsForLineno[1]{%
   \patchAmsMathEnvironmentForLineno{#1}%
   \patchAmsMathEnvironmentForLineno{#1*}}%
\AtBeginDocument{%
\patchBothAmsMathEnvironmentsForLineno{equation}%
\patchBothAmsMathEnvironmentsForLineno{align}%
\patchBothAmsMathEnvironmentsForLineno{flalign}%
\patchBothAmsMathEnvironmentsForLineno{alignat}%
\patchBothAmsMathEnvironmentsForLineno{gather}%
\patchBothAmsMathEnvironmentsForLineno{multline}%
}

\newcommand{\pt}{\mbox{$p_T$}\xspace}

\newcommand{\sqsn}{\mbox{$\sqrt{s_{_{NN}}}$}\xspace}
\newcommand{\dau}{\mbox{$d$$+$Au}\xspace}
\newcommand{\pau}{\mbox{$p$$+$Au}\xspace}

\newcommand{\hau}{\mbox{$^3$He$+$Au}\xspace}
\newcommand{\pp}{\mbox{$p$$+$$p$}\xspace}

\newcommand{\pdhau}{\mbox{$p/d/^{3}$He$+$Au}\xspace}
\newcommand{\dndeta}{\mbox{$dN_{\rm ch}/d\eta$}\xspace}

\newcommand{\ampt}{{\sc ampt}\xspace}
\newcommand{\sonic}{{\sc sonic}\xspace}

\begin{document}

\title{Creating small circular, elliptical, and triangular droplets of quark-gluon plasma}


\newcommand{\abilene}{Abilene Christian University, Abilene, Texas 79699, USA}
\newcommand{\augie}{Department of Physics, Augustana University, Sioux Falls, South Dakota 57197, USA}
\newcommand{\banaras}{Department of Physics, Banaras Hindu University, Varanasi 221005, India}
\newcommand{\barc}{Bhabha Atomic Research Centre, Bombay 400 085, India}
\newcommand{\baruch}{Baruch College, City University of New York, New York, New York, 10010 USA}
\newcommand{\bnlcoll}{Collider-Accelerator Department, Brookhaven National Laboratory, Upton, New York 11973-5000, USA}
\newcommand{\bnlphys}{Physics Department, Brookhaven National Laboratory, Upton, New York 11973-5000, USA}
\newcommand{\caucr}{University of California-Riverside, Riverside, California 92521, USA}
\newcommand{\charlesczech}{Charles University, Ovocn\'{y} trh 5, Praha 1, 116 36, Prague, Czech Republic}
\newcommand{\chonbuk}{Chonbuk National University, Jeonju, 561-756, Korea}
\newcommand{\cns}{Center for Nuclear Study, Graduate School of Science, University of Tokyo, 7-3-1 Hongo, Bunkyo, Tokyo 113-0033, Japan}
\newcommand{\colorado}{University of Colorado, Boulder, Colorado 80309, USA}
\newcommand{\columbia}{Columbia University, New York, New York 10027 and Nevis Laboratories, Irvington, New York 10533, USA}
\newcommand{\czechtech}{Czech Technical University, Zikova 4, 166 36 Prague 6, Czech Republic}
\newcommand{\debrecen}{Debrecen University, H-4010 Debrecen, Egyetem t{\'e}r 1, Hungary}
\newcommand{\elte}{ELTE, E{\"o}tv{\"o}s Lor{\'a}nd University, H-1117 Budapest, P{\'a}zm{\'a}ny P.~s.~1/A, Hungary}
\newcommand{\eszterhazy}{Eszterh\'azy K\'aroly University, K\'aroly R\'obert Campus, H-3200 Gy\"ongy\"os, M\'atrai \'ut 36, Hungary}
\newcommand{\ewha}{Ewha Womans University, Seoul 120-750, Korea}
\newcommand{\fsu}{Florida State University, Tallahassee, Florida 32306, USA}
\newcommand{\gsu}{Georgia State University, Atlanta, Georgia 30303, USA}
\newcommand{\hiroshima}{Hiroshima University, Kagamiyama, Higashi-Hiroshima 739-8526, Japan}
\newcommand{\howard}{Department of Physics and Astronomy, Howard University, Washington, DC 20059, USA}
\newcommand{\ihepprot}{IHEP Protvino, State Research Center of Russian Federation, Institute for High Energy Physics, Protvino, 142281, Russia}
\newcommand{\illuiuc}{University of Illinois at Urbana-Champaign, Urbana, Illinois 61801, USA}
\newcommand{\inrras}{Institute for Nuclear Research of the Russian Academy of Sciences, prospekt 60-letiya Oktyabrya 7a, Moscow 117312, Russia}
\newcommand{\instpasczech}{Institute of Physics, Academy of Sciences of the Czech Republic, Na Slovance 2, 182 21 Prague 8, Czech Republic}
\newcommand{\isu}{Iowa State University, Ames, Iowa 50011, USA}
\newcommand{\jaea}{Advanced Science Research Center, Japan Atomic Energy Agency, 2-4 Shirakata Shirane, Tokai-mura, Naka-gun, Ibaraki-ken 319-1195, Japan}
\newcommand{\jyvaskyla}{Helsinki Institute of Physics and University of Jyv{\"a}skyl{\"a}, P.O.Box 35, FI-40014 Jyv{\"a}skyl{\"a}, Finland}
\newcommand{\kek}{KEK, High Energy Accelerator Research Organization, Tsukuba, Ibaraki 305-0801, Japan}
\newcommand{\korea}{Korea University, Seoul, 136-701, Korea}
\newcommand{\kurchatov}{National Research Center ``Kurchatov Institute", Moscow, 123098 Russia}
\newcommand{\kyoto}{Kyoto University, Kyoto 606-8502, Japan}
\newcommand{\lawllnl}{Lawrence Livermore National Laboratory, Livermore, California 94550, USA}
\newcommand{\losalamos}{Los Alamos National Laboratory, Los Alamos, New Mexico 87545, USA}
\newcommand{\lund}{Department of Physics, Lund University, Box 118, SE-221 00 Lund, Sweden}
\newcommand{\lyon}{IPNL, CNRS/IN2P3, Univ Lyon, Université Lyon 1, F-69622, Villeurbanne, France}
\newcommand{\maryland}{University of Maryland, College Park, Maryland 20742, USA}
\newcommand{\mass}{Department of Physics, University of Massachusetts, Amherst, Massachusetts 01003-9337, USA}
\newcommand{\michigan}{Department of Physics, University of Michigan, Ann Arbor, Michigan 48109-1040, USA}
\newcommand{\muhlenberg}{Muhlenberg College, Allentown, Pennsylvania 18104-5586, USA}
\newcommand{\nara}{Nara Women's University, Kita-uoya Nishi-machi Nara 630-8506, Japan}
\newcommand{\natmephi}{National Research Nuclear University, MEPhI, Moscow Engineering Physics Institute, Moscow, 115409, Russia}
\newcommand{\newmex}{University of New Mexico, Albuquerque, New Mexico 87131, USA}
\newcommand{\nmsu}{New Mexico State University, Las Cruces, New Mexico 88003, USA}
\newcommand{\ohio}{Department of Physics and Astronomy, Ohio University, Athens, Ohio 45701, USA}
\newcommand{\ornl}{Oak Ridge National Laboratory, Oak Ridge, Tennessee 37831, USA}
\newcommand{\orsay}{IPN-Orsay, Univ.~Paris-Sud, CNRS/IN2P3, Universit\'e Paris-Saclay, BP1, F-91406, Orsay, France}
\newcommand{\peking}{Peking University, Beijing 100871, People's Republic of China}
\newcommand{\pnpi}{PNPI, Petersburg Nuclear Physics Institute, Gatchina, Leningrad region, 188300, Russia}
\newcommand{\riken}{RIKEN Nishina Center for Accelerator-Based Science, Wako, Saitama 351-0198, Japan}
\newcommand{\rikjrbrc}{RIKEN BNL Research Center, Brookhaven National Laboratory, Upton, New York 11973-5000, USA}
\newcommand{\rikkyo}{Physics Department, Rikkyo University, 3-34-1 Nishi-Ikebukuro, Toshima, Tokyo 171-8501, Japan}
\newcommand{\saispbstu}{Saint Petersburg State Polytechnic University, St.~Petersburg, 195251 Russia}
\newcommand{\seoulnat}{Department of Physics and Astronomy, Seoul National University, Seoul 151-742, Korea}
\newcommand{\stonybrkc}{Chemistry Department, Stony Brook University, SUNY, Stony Brook, New York 11794-3400, USA}
\newcommand{\stonycrkp}{Department of Physics and Astronomy, Stony Brook University, SUNY, Stony Brook, New York 11794-3800, USA}
\newcommand{\tenn}{University of Tennessee, Knoxville, Tennessee 37996, USA}
\newcommand{\titech}{Department of Physics, Tokyo Institute of Technology, Oh-okayama, Meguro, Tokyo 152-8551, Japan}
\newcommand{\tsukuba}{Tomonaga Center for the History of the Universe, University of Tsukuba, Tsukuba, Ibaraki 305, Japan}
\newcommand{\vandy}{Vanderbilt University, Nashville, Tennessee 37235, USA}
\newcommand{\weizmann}{Weizmann Institute, Rehovot 76100, Israel}
\newcommand{\wigner}{Institute for Particle and Nuclear Physics, Wigner Research Centre for Physics, Hungarian Academy of Sciences (Wigner RCP, RMKI) H-1525 Budapest 114, POBox 49, Budapest, Hungary}
\newcommand{\yonsei}{Yonsei University, IPAP, Seoul 120-749, Korea}
\newcommand{\zagreb}{Department of Physics, Faculty of Science, University of Zagreb, Bijeni\v{c}ka c.~32 HR-10002 Zagreb, Croatia}
\affiliation{\abilene}
\affiliation{\augie}
\affiliation{\banaras}
\affiliation{\barc}
\affiliation{\baruch}
\affiliation{\bnlcoll}
\affiliation{\bnlphys}
\affiliation{\caucr}
\affiliation{\charlesczech}
\affiliation{\chonbuk}
\affiliation{\cns}
\affiliation{\colorado}
\affiliation{\columbia}
\affiliation{\czechtech}
\affiliation{\debrecen}
\affiliation{\elte}
\affiliation{\eszterhazy}
\affiliation{\ewha}
\affiliation{\fsu}
\affiliation{\gsu}
\affiliation{\hiroshima}
\affiliation{\howard}
\affiliation{\ihepprot}
\affiliation{\illuiuc}
\affiliation{\inrras}
\affiliation{\instpasczech}
\affiliation{\isu}
\affiliation{\jaea}
\affiliation{\jyvaskyla}
\affiliation{\kek}
\affiliation{\korea}
\affiliation{\kurchatov}
\affiliation{\kyoto}
\affiliation{\lawllnl}
\affiliation{\losalamos}
\affiliation{\lund}
\affiliation{\lyon}
\affiliation{\maryland}
\affiliation{\mass}
\affiliation{\michigan}
\affiliation{\muhlenberg}
\affiliation{\nara}
\affiliation{\natmephi}
\affiliation{\newmex}
\affiliation{\nmsu}
\affiliation{\ohio}
\affiliation{\ornl}
\affiliation{\orsay}
\affiliation{\peking}
\affiliation{\pnpi}
\affiliation{\riken}
\affiliation{\rikjrbrc}
\affiliation{\rikkyo}
\affiliation{\saispbstu}
\affiliation{\seoulnat}
\affiliation{\stonybrkc}
\affiliation{\stonycrkp}
\affiliation{\tenn}
\affiliation{\titech}
\affiliation{\tsukuba}
\affiliation{\vandy}
\affiliation{\weizmann}
\affiliation{\wigner}
\affiliation{\yonsei}
\affiliation{\zagreb}
\author{C.~Aidala} \affiliation{\michigan} 
\author{Y.~Akiba} \email[PHENIX Spokesperson: ]{akiba@rcf.rhic.bnl.gov} \affiliation{\riken} \affiliation{\rikjrbrc} 
\author{M.~Alfred} \affiliation{\howard} 
\author{V.~Andrieux} \affiliation{\michigan} 
\author{K.~Aoki} \affiliation{\kek} 
\author{N.~Apadula} \affiliation{\isu} 
\author{H.~Asano} \affiliation{\kyoto} \affiliation{\riken} 
\author{C.~Ayuso} \affiliation{\michigan} 
\author{B.~Azmoun} \affiliation{\bnlphys} 
\author{V.~Babintsev} \affiliation{\ihepprot} 
% ...
\collaboration{PHENIX Collaboration}  \noaffiliation

\maketitle
\textbf{ The experimental study of the collisions of heavy nuclei at
relativistic energies has established the properties of the quark-gluon plasma
(QGP), a state of hot, dense nuclear matter in which quarks and gluons are not
bound into hadrons~\cite{Arsene:2004fa,Back:2004je,Adams:2005dq,Adcox:2004mh}.
In this state, matter behaves as a nearly inviscid fluid~\cite{Heinz:2013th}
that efficiently translates initial spatial anisotropies into correlated
momentum anisotropies among the produced particles, creating a common velocity 
field pattern known as collective flow. In
recent years, comparable momentum anisotropies have been measured in 
small-system proton-proton ($p$$+$$p$) and proton-nucleus ($p$$+$$A$) collisions, despite
expectations that the volume and lifetime of the medium produced would be too
small to form a QGP. Here, we report on the observation of elliptic and
triangular flow patterns of charged particles produced in proton-gold ($p$$+$Au),
deuteron-gold ($d$$+$Au), and helium-gold ($^3$He$+$Au)  collisions at a nucleon-nucleon
center-of-mass energy $\sqrt{s_{_{NN}}}$~=~200 GeV. The unique combination of three distinct
initial geometries and two flow patterns provides unprecedented model
discrimination. Hydrodynamical models, which include the formation of a 
short-lived QGP droplet, provide a simultaneous description of these 
measurements. }


Experiments at the Relativistic Heavy Ion Collider (RHIC) and the Large Hadron 
Collider (LHC) explore emergent phenomena in quantum chromodynamics, most 
notably the near-perfect fluidity of the QGP. To quantify this behavior, 
the azimuthal distribution of each event's final-state particles, $\frac{dN}{d\phi}$, is 
decomposed into a Fourier series as follows:
\begin{linenomath}
 \begin{equation}
   \frac{dN}{d\phi} \propto 1+ \sum_n 2 v_n(p_T)\cos(n(\phi-\psi_n)),
   \label{eq:dndphi}
 \end{equation}
\end{linenomath}
where \pt and $\phi$ are the transverse momentum and the azimuthal angle of a 
particle relative to the beam direction, respectively, and $\psi_n$ is the orientation of the $n^{\rm th}$ 
order symmetry plane of the produced particles.  The second ($v_2$) and third ($v_3$) Fourier coefficients 
represent the amplitude of elliptic and triangular flow, respectively.  
A multitude of measurements of the Fourier coefficients, utilizing a variety of 
techniques, have been well-described by hydrodynamical models, thereby 
establishing the fluid nature of the QGP in large-ion 
collisions~\cite{Heinz:2013th}. 

The LHC experiments were first to observe  similar features in small-system
collisions~\cite{Khachatryan:2010gv,CMS:2012qk,Abelev:2012ola,Aad:2012gla},  followed
closely by reanalysis of previously recorded \dau data from
RHIC~\cite{Adare:2013piz,Adare:2014keg}. These unexpected results highlighted the need to
explore whether these smallest hadronic systems still form QGP. Alternatively, 
a number of physics mechanisms that do not involve QGP formation have been proposed, 
including those which attribute final-state momentum anisotropy to momentum
correlations generated at the earliest stages of the collision, hence referred to as initial-state
momentum correlation models (see Refs.~\cite{Dusling:2015gta}
and~\cite{Nagle:2018nvi} for recent reviews).

\begin{figure*}[htbp]
  \begin{center}
    \includegraphics[width=1.0\linewidth]{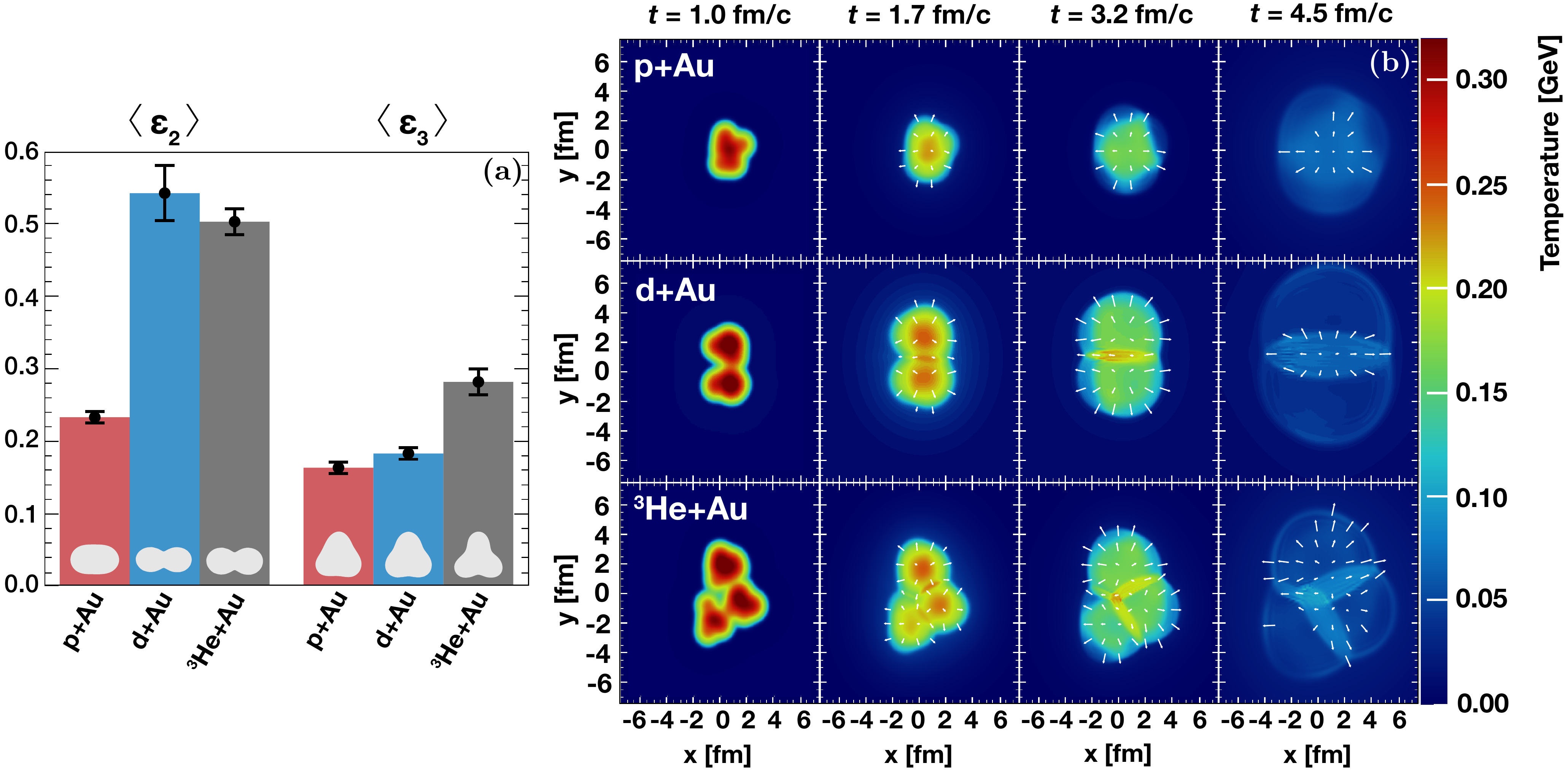}
    \caption{$\mid$ {\bf Average system 
    		eccentricities from a Monte Carlo 
            Glauber model and hydrodynamic 
            evolution of small systems.}  
            {\bf a}, Average second (third) order 
            spatial eccentricities, $\varepsilon_2$ 
            ($\varepsilon_3$), shown as columns 
            for small impact parameter \pau (red), 
            \dau (blue), and \hau (black) collisions 
            as calculated from a MC Glauber model. 
            The second and third order spatial
             eccentricities correspond to ellipticity 
             and triangularity respectively as depicted 
             by the shapes inset in the bars. 
    		 {\bf b}, Hydrodynamic evolution of a 
             characteristic head-on \pau (top), 
             \dau (middle), and \hau (bottom) collision 
             at \sqsn~=~200 GeV as calculated by \sonic, 
             where the $p/d/^3$He completely overlap 
             with the Au nucleus. 
             From left to right each row gives the 
             temperature distribution of 
             the nuclear matter at four time 
             points following the initial 
             collision at $t=0$. The arrows depict 
             the velocity field, with the length of 
             the longest arrow plotted 
             corresponding to $\beta=0.82$.}
    \label{fig:hydro}
  \end{center}
\end{figure*}

A projectile geometry scan utilizing the unique capabilities of RHIC was 
proposed in Ref.~\cite{Nagle:2013lja} in order to discriminate between hydrodynamical models 
that couple to the initial geometry and initial-state momentum correlation models that do not. 
Varying the collision system from \pau, to \dau, to \hau changes the initial 
geometry from dominantly circular, to elliptical, and to triangular 
configurations, respectively, as characterized by the 
2$^{\rm nd}$ and 3$^{\rm rd}$ order spatial eccentricities, which correspond to 
ellipticity and triangularity, respectively. The $n^{\rm th}$ order spatial 
eccentricity of the system, $\varepsilon_n$, typically determined from a Monte Carlo (MC) Glauber 
model of nucleon-nucleon interactions (see e.g. Ref~\cite{Miller:2007ri}), can 
be defined as
\begin{linenomath}
  \begin{equation}
    \varepsilon_n=\frac{\sqrt{\langle r^n\cos(n\phi)\rangle^2+\langle r^n \sin
    (n\phi)\rangle^2}}{\langle r^n \rangle},
    \label{eq:epsilon_n}
  \end{equation}
\end{linenomath}
where $r$ and $\phi$ are the polar coordinates of participating 
nucleons~\cite{Alver:2010gr}. The eccentricity fluctuates event-by-event and is 
generally dependent on the impact parameter of the collision and the number of 
participating nucleons. The mean $\varepsilon _2$ and $\varepsilon _3$ values 
for small impact parameter \pdhau collisions are shown in Fig.~\ref{fig:hydro}a.
The $\varepsilon _2$ and $\varepsilon _3$ values in \dau and \hau are 
driven almost entirely by the intrinsic geometry of the deuteron and $^3$He, 
while the values in \pau collisions are driven by fluctuations in the configuration
of struck nucleons in the Au nucleus, as the proton itself is on average
circular. 

%
%

Hydrodynamical models begin with an initial spatial energy-density distribution with a given temperature that evolves in time following the laws of
relativistic viscous hydrodynamics using an equation of state determined from lattice QCD~\cite{Gale:2013da}. 
Examples of this evolution are shown for
\pdhau collisions in Fig.~\ref{fig:hydro}b using the hydrodynamical model 
\sonic~\cite{Habich:2014jna}. The first panel of each row shows
the temperature profile at time $t=1.0$ fm/$c$ for typical \pau , \dau, and 
\hau collisions.  The following three panels show
snapshots of the temperature evolution at three different time points. The
initial spatial distribution also sets the pressure gradient field, which translates
into a velocity field, which in turn determines the 
azimuthal momentum distribution of produced particles. The relative magnitude and direction 
of the velocity is represented in the figure by arrows. At
the final time point, $t=4.5$ fm/c, the mostly circular (top), elliptical
(middle), and triangular (bottom) initial spatial eccentricities have been
translated into dominantly radial, elliptic, and triangular flow, respectively.
Given these different initial geometries, as characterized by the 
$\varepsilon_2$ and $\varepsilon_3$ values shown in Fig.~\ref{fig:hydro}a, 
hydrodynamical models provide a clear prediction for the ordering of the 
experimentally accessible $v_2$ and $v_3$ signals, following that of the 
$\varepsilon_n$, namely
\begin{linenomath}
  \begin{equation}
    \begin{aligned}
      v_2^{\pau}<v_2^{\dau}\approx v_2^{\hau}, \\
      v_3^{\pau}\approx v_3^{\dau}<v_3^{\hau}.
      \label{eq:vn-hydro}
    \end{aligned}
  \end{equation}
\end{linenomath}
This ordering assumes that hydrodynamics can efficiently translate the initial 
geometric $\varepsilon_n$ into dynamical $v_n$, which in turn requires a 
small value for the specific shear viscosity.


There exist a class of alternative explanations where $v_{n}$ is not generated via flow, but rather
is created at the earliest time in the collision process as described by
so-called initial-state momentum correlation models.  They produce a mimic flow signal 
where the initial collision generates color flux tubes that have a preference to emit
particles back-to-back in azimuth~\cite{Dusling:2012iga,Dumitru:2010iy}.   These 
color flux tubes, also referred to as domains, have a transverse size relative to the collision axis less than 
the color-correlation length of order 0.1-0.2~fm.   In the case where individual domains are resolved, a collision system with a larger overall area but the same characteristic 
domain size (for example \dau and \hau compared with \pau and \pp)
should have a weaker correlation because the different domains are separated
and do not communicate~\cite{Dumitru:2014dra,Lappi:2015vta}.  An instructive analogy is a ferromagnet with many domains: if the domains are
separated and disconnected, the overall magnetic field is weakened by the
cancellation of effects from the random orientation in the different domains. The RMS diameter of the deuteron is
4.2~fm, and so in \dau collisions the two hot spots are much further apart than the 
characteristic domain size.
A straightforward prediction is then that the $v_{2}$ and  $v_{3}$ coefficients should be ordered
\begin{linenomath}
  \begin{equation}
    \begin{aligned}
      v_n^{\pau}>v_n^{\dau}>v_n^{\hau},
      \label{eq:vn-momentum}
    \end{aligned}
  \end{equation}
\end{linenomath}
in contradistinction to the hydrodynamic flow prediction.


An experimental realization of the proposed geometry scan has been under way
since 2014 at RHIC. Collisions of \hau, \pau, and \dau at \sqsn = 200 GeV were recorded in 2014,
2015, and 2016, respectively.   The PHENIX experiment observed elliptic
anisotropies in the azimuthal distributions of the charged particles produced
in all three systems~\cite{Aidala:2016vgl,Aidala:2017pup,Adare:2015ctn}, as well
as triangular anisotropies in \hau collisions~\cite{Adare:2015ctn}. This Letter
completes this set of elliptic and triangular flow measurements from
PHENIX in all three systems and explores the relation between the strength of
the measured $v_n$ and the initial-state geometry.


The $v_n$ measurements reported here are determined using the event plane
method~\cite{Voloshin:1994mz} for charged hadrons in the midrapidity region
covering $|\eta|<0.35$, where $\eta$ is the particle 
pseudorapidity,  
\begin{linenomath}
  \begin{equation}
    \begin{aligned}
      \eta \equiv -\ln\left(\tan\frac{\theta}{2}\right),
      \label{eq:eta}
    \end{aligned}
  \end{equation}
\end{linenomath}
and $\theta$ is the polar angle of the particle. 
The $2^{\rm nd}$ order event plane is determined using detectors in the Au-going direction covering $-3.0<\eta<-1.0$ in $p/d$$+$Au and $-3.9<\eta<-3.1$ in \hau. The $3^{\rm rd}$ order event plane is determined using detectors in the
Au-going direction covering $-3.9<\eta<-3.1$ in all cases. The pseudorapidity gap between the
particle measurements and the event plane determination 
excludes auto-correlations and reduces short-range correlations arising from, for
example, jets and particle decays---typically referred to as nonflow correlations. Estimates
of possible remaining nonflow contributions are included in the systematic
uncertainties. Additional uncertainties related to detector alignment, data
selection, and event plane determination are also included in the systematic
uncertainty estimation. In these small collision systems
the event plane resolution is low, meaning that $v_n\{\textrm{EP}\}=\sqrt{\langle v_n^2\rangle}$~\cite{Ollitrault:2009it}
and the results are therefore equivalent to measurements using two-particle 
correlation methods.

\begin{figure}[htp]
  \begin{center}
    \includegraphics[width=0.8\linewidth]{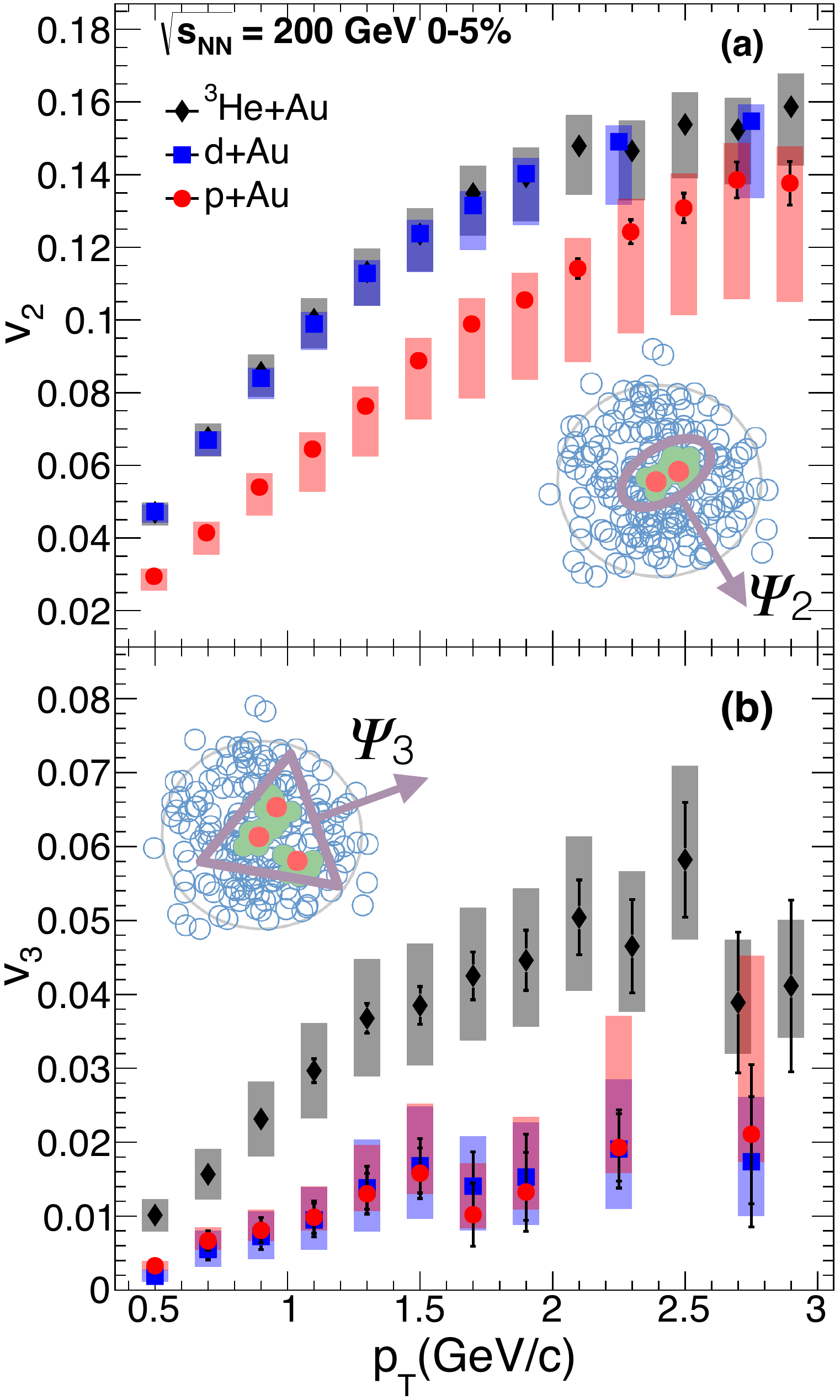}
    \caption{$\mid$ {\bf Measured $v_n(\pt)$ in three collision systems.} 
    		 {\bf a}, Measurements of $v_2(\pt)$ in the 0-5\% most central \pau, \dau, and \hau collisions at \sqsn~=~200 GeV. A \dau event from a MC 
             Glauber model
             is inset with the elliptic symmetry plane angle, $\psi_2$, 
             depicted. {\bf b}, Measurements of $v_3(\pt)$ in the 0-5\% most central \pau, \dau, and \hau collisions at \sqsn~=~200 GeV. A \hau event
             from a MC Glauber model
             is inset with the triangular symmetry plane angle, $\psi_3$, 
             depicted. Each point in {\bf a,b} represents an average over 
             $p_T$ bins of width 0.2 GeV/c to 0.5 GeV/c; black diamonds 
             are \hau, blue squares are \dau, red circles are \pau. 
             Line error bars are statistical and box error bars are 
             systematic (Methods).}
    \label{fig:vn-data}
  \end{center}
\end{figure}

Measurements of $v_n$ as a function of \pt are shown for all three systems in
Fig.~\ref{fig:vn-data}. The measurements are performed in the 0-5\% most central events, 
an experimentally determined criterion which selects the 5\% of events with the
largest number of produced particles (hereafter referred to simply as
``multiplicity'') in the region $-3.9<\eta<-3.1$.  A detailed description of the
centrality determination in small systems is given in Ref.~\cite{Adare:2013nff}.
The vertical bars on each point represent the
statistical uncertainties, while the shaded  boxes represent the systematic uncertainties. 
The flow  coefficients follow the
prediction of hydrodynamical models shown in equation (\ref{eq:vn-hydro}).
These relationships suggest that the primary  driver of azimuthal
momentum anisotropies in particle emission is initial  spatial anisotropy.

%
\begin{figure*}[htbp]
  \begin{center}
    \includegraphics[width=1.0\linewidth]{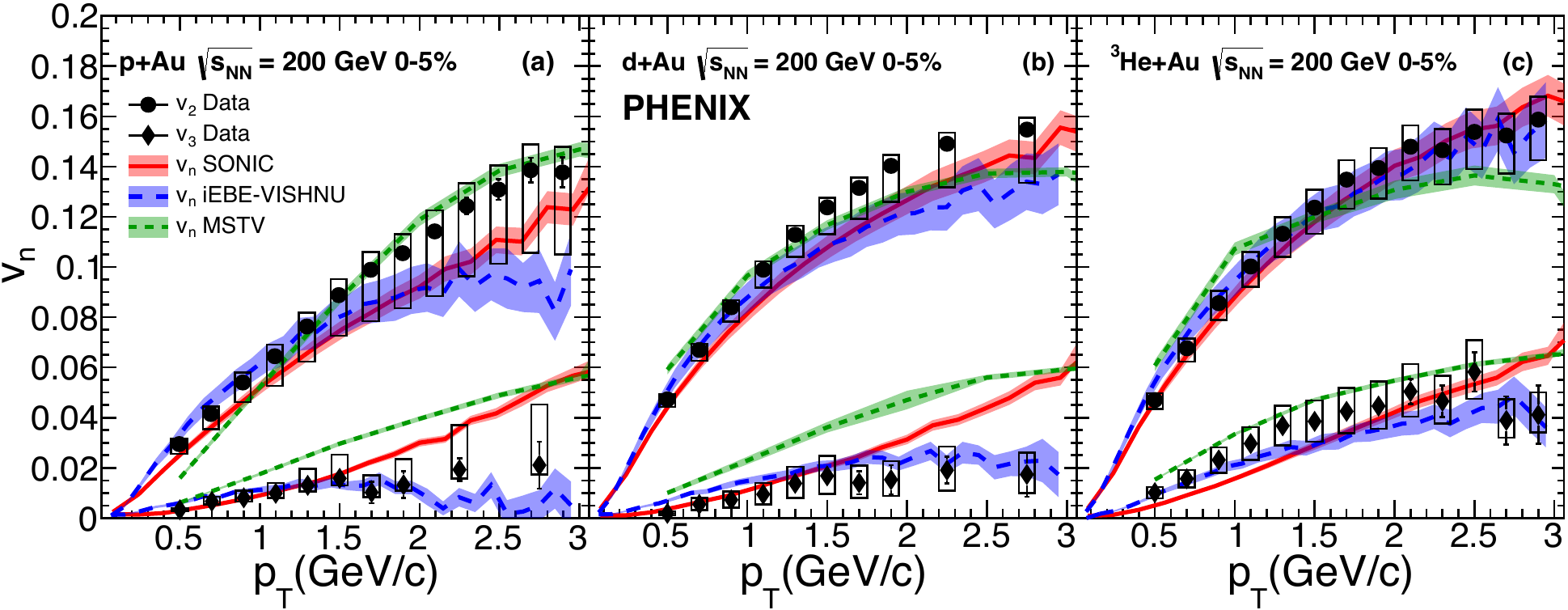}
    \caption{$\mid$ {\bf Measured $v_n(\pt)$ in three collision systems
             compared to models.} {\bf a}, Measured $v_n(\pt)$
             in the 0-5\% most central \pau collisions compared to models. {\bf b}, Measured
             $v_n(\pt)$ in the 0-5\% most central \dau collisions compared to models. {\bf c}, 
             Measured $v_n(\pt)$ in the 0-5\% most central \hau compared to models. Each
             point in {\bf a-c} represents an average over $p_T$ bins of width
             0.2 GeV/c to 0.5 GeV/c; black circles are $v_2$, black diamonds are 
             $v_3$. The solid red (dashed blue) curves in {\bf a-c} represent hydrodynamic predictions of $v_n$ from \sonic (iEBE-VISHNU). The solid green curves in {\bf a-c} represent initial-state momentum correlation postdictions of $v_n$ from MSTV.}
    \label{fig:vn-models}
  \end{center}
\end{figure*}

While Fig.~\ref{fig:vn-data} offers qualitative support for the hydrodynamic
theory, Fig.~\ref{fig:vn-models} directly compares these data to predictions
from two hydrodynamical models, \sonic~\cite{Habich:2014jna} (used in
Fig.~\ref{fig:hydro}) and iEBE-VISHNU~\cite{Shen:2016zpp}. The core structure of 
the two models is similar: the initial conditions are
evolved using viscous hydrodynamics, the fluid hadronizes, 
hadronic scattering occurs, and the $v_n$ coefficients of the final-state hadron distributions are determined using two-particle correlation methods. However, the detailed implementations are different, 
including the use of different fluctuations in the initial 
energy deposited, as well as different hadronic rescattering packages. Both 
calculations in Fig.~\ref{fig:vn-models} use a ratio of the shear viscosity $\eta$ to entropy density $s$ 
of $\eta/s=0.08\approx\frac{1}{4\pi}$, the conjectured lower limit in strongly-coupled field theories~\cite{Kovtun:2004de}.

Figure~\ref{fig:vn-models} shows that the hydrodynamical models are consistent with the $v_2$
data in all three systems. Both models capture the magnitude
difference of $v_3$ compared to $v_2$, the collision system dependence, as well 
as the general \pt dependence of $v_3$. The models tend to diverge at higher \pt in the case of $v_3$, which may be more sensitive to the hadronic
rescattering. To quantify the agreement, we calculate $p$-values following the procedure of incorporating data systematic uncertainties and their correlations into a modified $\chi^2$ analysis laid out in Ref.~\cite{Adare:2008cg} (See \textit{Methods} for details). We find that \sonic and iEBE-VISHNU yield combined $p$-values across the 6 measurements of 0.96 and 0.061 respectively. The large difference in $p$-values is driven by the effect of the dominant nonflow uncertainty, which is asymmetric and anti-correlated between $v_2$ and $v_3$. \sonic tends to underestimate the $v_2$ and overestimate the $v_3$, particularly in \pau and \dau, which is more in line with the uncertainty correlations than iEBE-VISHNU, which tends to yield a poorer description of the $p_T$ slope. Overall, the simultaneous description of these two observables in three different systems using a common initial geometry model and the same specific $\eta/s$ strongly supports the hydrodynamic
picture. 

The hydrodynamic calculations shown in Fig.~\ref{fig:vn-models} use initial conditions generated from a nucleon Glauber model. However, initial geometries with quark substructure do not significantly change the $\varepsilon_2$ and $\varepsilon_3$ values for high multiplicity \pdhau collisions~\cite{Welsh:2016siu,Weller:2017tsr} and thus the hydrodynamic results should be relatively insensitive to these variations.


While we have focused on hydrodynamical models here, there
is an alternative class of models that also translate initial spatial eccentricity to
final state particle azimuthal momentum anisotropy. Instead of hydrodynamic
evolution, the translation occurs via parton-parton scattering with a modest
interaction cross section.  These parton transport models, for example A
Multi-Phase Transport (\ampt) Model~\cite{Lin:2004en}, are able to capture the 
system ordering of $v_n$ at low-\pt in small systems~\cite{Koop:2015wea}, but 
fail to describe the \pt dependence and overall magnitude of the coefficients 
for all systems resulting in a $p$-value consistent with zero when compared to 
the data shown here. We have additionally analyzed \ampt following the identical 
PHENIX event plane method and find even worse agreement with the experimental data.

While the initial geometry models for the \dau and \hau are largely constrained
by our detailed understanding of the 2- and 3-body nucleon correlations in the
deuteron and $^3$He nuclei, respectively, the distribution of deposited energy around
each nucleon-nucleon collision site could result in an ambiguity between the
allowed ranges of the $\eta/s$ and the broadening of the initial
distribution, as pointed out in Ref.~\cite{Nagle:2018nvi}.  However, a broader
distribution of deposited energy results in a significant reduction of the
$\varepsilon_2$ values and an even greater reduction of $\varepsilon_3$, with by
far the largest reduction in the \pau system.  Here again, the simultaneous
constraints of the elliptic and triangular flow ordering eliminates this
ambiguity.

\begin{figure}
	\centering
	\includegraphics[width=0.98\columnwidth]{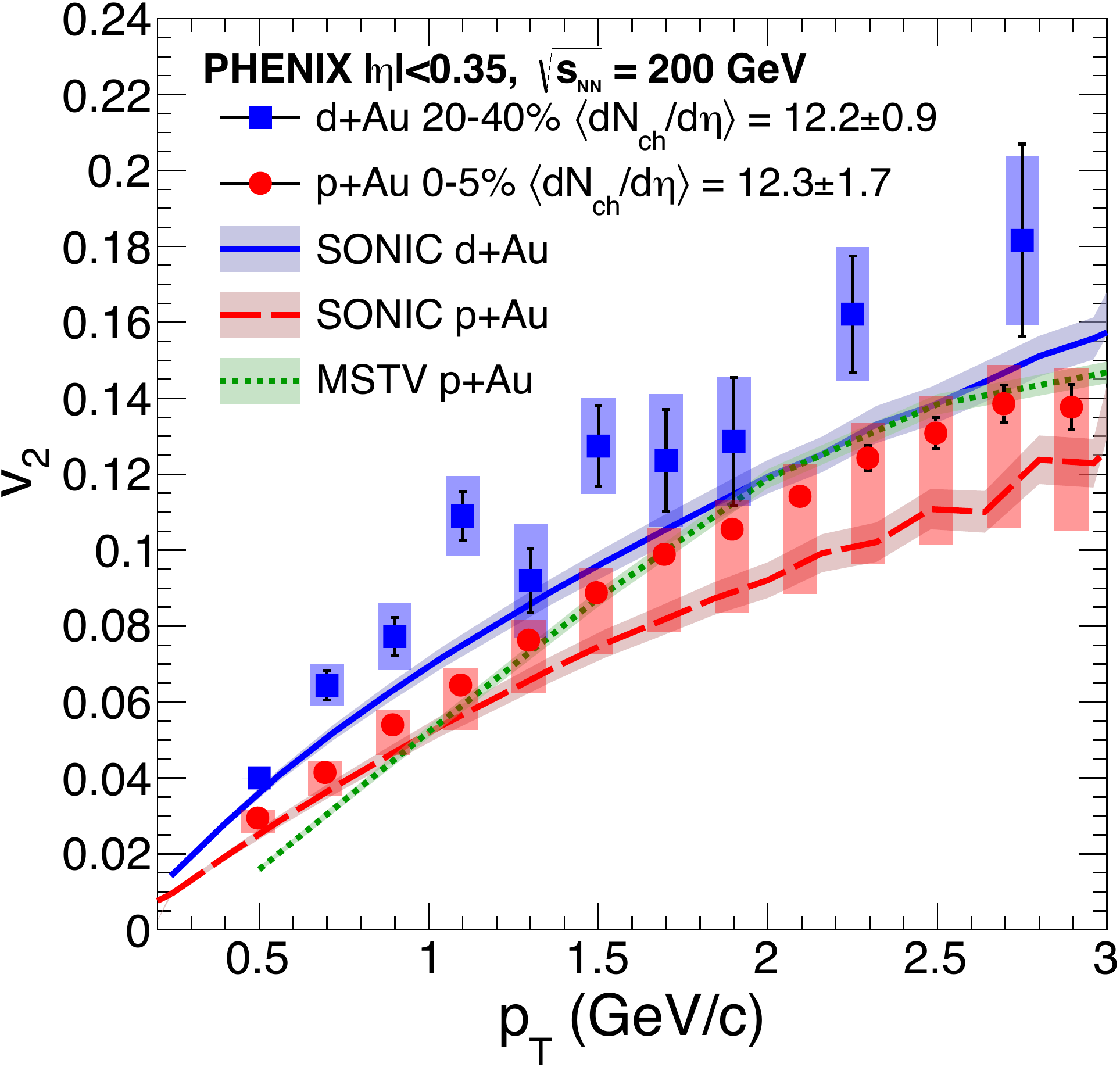}
	\caption{$\mid$ {\bf Measured $v_2(\pt)$ in \pau and \dau collisions at the same event multiplicity.} Measured $v_2(\pt)$
             in the 0-5\% most central \pau collisions and 20-40\% central \dau collisions compared to \sonic predictions and MSTV post-dictions. 
             Each point represents an average over 
             $p_T$ bins of width 0.2 GeV/c to 0.5 GeV/c;  blue circles are \dau, red circles are \pau. 
             Line error bars are statistical and box error bars are 
             systematic (Methods). The quoted \dndeta values are taken from Ref.~\cite{Adare:2018toe}. Blue and red curves correspond to \sonic predictions for \dau and \pau, respectively. The green curve corresponds to MSTV calculations for 0-5\% central \pau collisions, which the authors state are also applicable to \dau collisions at the same multiplicity.}
    \label{fig:samemult}
\end{figure}

Our experimental data also rule out the initial-state correlations scenario 
where color domains are individually resolved as the dominant mechanics for 
creating $v_2$ and $v_3$ in \pdhau collisions.  After our results became publicly 
available, a new calculation was presented in Ref.~\cite{Mace:2018vwq}, hereafter
referred to as MSTV, where the ordering of the
measured $v_n$ values matches the experimental data.  This calculation posits that gluons from the Au target do not
resolve individual color domains in the projectile $p/d/^3$He and interact with them
coherently, and thus the ordering does not follow Eq.~\ref{eq:vn-momentum}.  
The calculations are shown in Fig.~\ref{fig:vn-models}, and yield a $p$-value for the 
MSTV calculations of $v_2$ and $v_3$ for the three collision systems of effectively
zero, in contradistinction to the robust values found for the hydrodynamic models.
Another key statement made by MSTV -- that in the dilute-dense limit the 
saturation scale $Q_s^2$ is proportional to the number of produced charged 
particles -- is questionable~\cite{Nagle:2018ybc}, but also leads the MSTV authors to make a clear prediction
that the $v_2$ will be identical between systems when selecting on the same event
multiplicity. Shown in Fig.~\ref{fig:samemult} are the previously published \dau (20-40\%)
and \pau (0-5\%) $v_2$ where the measured mean charged particle multiplicities (\dndeta)
match~\cite{Adare:2018toe}. The results do not support the MSTV prediction of an 
identical $v_2$ for these two systems at the same multiplicity, while the 
differences in $v_2$ between the systems follow the expectations from 
hydrodynamic calculations matched to the same \dndeta.


In summary, we have shown azimuthal particle correlations in three different
small-system collisions with different intrinsic initial geometries. The
simultaneous constraints of $v_2$ and $v_3$ in $p/d/^3$He$+$Au collisions
definitively demonstrate that the $v_n$'s are correlated to the initial
geometry, removing any ambiguity related to event multiplicity or initial
geometry models. We find that initial-state momentum correlation models where 
color domains are individually resolved are ruled out as the dominant mechanism
behind the observed collectivity. New calculations where the
domains are not resolved are unable to simultaneously explain the $v_2$ and $v_3$ 
in high multiplicity collisions, and are further unable to explain the 
difference in $v_2$ between \pau and \dau when the multiplicity selections are
matched. Further, we find that hydrodynamical models which include
QGP formation provide a simultaneous and quantitative description of the data in
all three systems.


\textbf{Acknowledgements} We thank the staff of the Collider-Accelerator and Physics
Departments at Brookhaven National Laboratory and the staff of
the other PHENIX participating institutions for their vital
contributions.  We acknowledge support from the 
Office of Nuclear Physics in the
Office of Science of the Department of Energy,
the National Science Foundation, 
Abilene Christian University Research Council, 
Research Foundation of SUNY, and
Dean of the College of Arts and Sciences, Vanderbilt University 
(U.S.A),
Ministry of Education, Culture, Sports, Science, and Technology
and the Japan Society for the Promotion of Science (Japan),
Conselho Nacional de Desenvolvimento Cient\'{\i}fico e
Tecnol{\'o}gico and Funda\c c{\~a}o de Amparo {\`a} Pesquisa do
Estado de S{\~a}o Paulo (Brazil),
Natural Science Foundation of China (People's Republic of China),
Croatian Science Foundation and
Ministry of Science and Education (Croatia),
Ministry of Education, Youth and Sports (Czech Republic),
Centre National de la Recherche Scientifique, Commissariat
{\`a} l'{\'E}nergie Atomique, and Institut National de Physique
Nucl{\'e}aire et de Physique des Particules (France),
Bundesministerium f\"ur Bildung und Forschung, Deutscher
Akademischer Austausch Dienst, and Alexander von Humboldt Stiftung (Germany),
NKFIH, EFOP, the New National Excellence Program ({\'U}NKP) and the J. Bolyai
Research Scholarships (Hungary),
Department of Atomic Energy and Department of Science and Technology (India), 
Israel Science Foundation (Israel), 
Basic Science Research Program through NRF of the Ministry of Education (Korea),
Physics Department, Lahore University of Management Sciences (Pakistan),
Ministry of Education and Science, Russian Academy of Sciences,
Federal Agency of Atomic Energy (Russia),
VR and Wallenberg Foundation (Sweden), 
the U.S. Civilian Research and Development Foundation for the
Independent States of the Former Soviet Union, 
the Hungarian American Enterprise Scholarship Fund,
and the US-Israel Binational Science Foundation.

\textbf{Author contributions} All PHENIX collaboration members contributed to
the publication of these results in a variety of roles including detector
construction, data collection, data processing, and analysis. A subset of
collaboration members prepared this manuscript, and all authors had the opportunity to review the final version.

\textbf{Competing financial interests} The authors declare no competing 
financial interests.

%

\textbf{Methods} Here we provide details of the $v_3$ measurements in \pau and
\dau collisions as well as details on quantifying comparisons of theory to data. 
For details on the remaining measurements see 
Refs.~\cite{Aidala:2016vgl,Aidala:2017pup,Adare:2015ctn}.

\textit{Experimental Setup:} These measurements utilize the PHENIX detector at
RHIC. Particle tracking is performed by two arms at midrapidity, each covering
$|\eta|<0.35$ and $\frac{\pi}{2}$ in azimuth using drift chambers (DC) and pad
chambers (PC)~\cite{Adcox:2003zp}. Beam-beam counters (BBC) located at forward
and backward rapidities ($3.1<|\eta|<3.9$), each consisting of an array of 64
quartz Cherenkov radiators read out by photomultiplier
tubes~\cite{Adare:2013nff}, provide event triggering, collision vertexing, and event
plane angle determination. Additionally, a forward vertex detector (FVTX)
covering $1.0<|\eta|<3.0$ and composed of high efficiency silicon 
mini-strips~\cite{Aidala:2013vna} provides an independent event plane angle determination.
A description of the PHENIX detector can be found in Ref.~\cite{Adcox:2003zm}.

\textit{Event Selection:}  A minimum bias (MB) interaction trigger is provided by
the BBC, which requires at least one hit tube in both the south ($\eta<0$, Au-going direction) and north ($\eta>0$, $p/d$-going direction), along
with an online vertex within $|z_{\rm vertex}|<10$ cm of the nominal interaction
region. In addition to the MB trigger, a high multiplicity trigger requiring
$>35$ ($>40$) hit tubes in the BBCS provided a factor of 25 (188)
enhancement of high multiplicity events in \pau (\dau) collisions.
A more precise offline collision vertex is
determined using timing information in the BBC and is constrained to
$|z_{\rm vertex}|<10$ cm in order to be sufficiently inside the acceptance of the
detector. Events containing more than one nucleus-nucleus collision, referred to
as double interaction events, are rejected using an algorithm based on BBC charge and timing
information described in Ref.~\cite{Aidala:2017pup}. Event centrality is
determined using the total charge collected in the south BBC, as described in
Ref.~\cite{Adare:2013nff}. We require an event centrality of 0--5\% to select
events with the highest multiplicity, where the signal of interest is strongest.
In total, 322 (636) million \pau (\dau) events are analyzed.


\textit{Track Selection:}  Quality cuts are applied to reconstructed particle
tracks requiring hits in both the DC and the outermost PC layer with a
required $3\sigma$ level of agreement. This  removes the majority of tracks that
do not originate from the primary collision. Further details can be found in
Refs.~\cite{Aidala:2016vgl,Aidala:2017pup,Adare:2015ctn}. %

\textit{Event Plane Determination:}  The third-order symmetry plane angle,
$\psi_3$ is measured using the south BBC via the standard method~\cite{Poskanzer:1998yz}. Namely,
\begin{linenomath}
  \begin{equation}
    \psi_n = \frac{1}{n}\arctan{\frac{\sum_i^N \sin{n\phi_i}}{\sum_i^N \cos{n\phi_i}}},
  \end{equation}
\end{linenomath}
where $N$ is the number of particles and $\phi_i$ is the azimuthal angle of each particle. The $\psi_3$
resolution, $R(\psi_3)$, is calculated using the three-subevent method which
correlates measurements in the south BBC, south FVTX, and central arms. The calculated resolutions are
6.7\% and 5.7\% in \pau and \dau collisions, respectively.

\textit{Determination of $v_3$:}
The $v_3$ values are measured using the event plane method~\cite{Poskanzer:1998yz,Voloshin:1994mz} as
\begin{linenomath}
  \begin{equation}
    v_3 = \frac{\langle \cos(3(\phi-\psi_3) \rangle}{R(\psi_3)},
    \label{eq:vn}
  \end{equation}
\end{linenomath}
where $\phi$ is the azimuthal angle of particles emitted at midrapitiy, $|\eta|<0.35$.

\textit{Systematic uncertainties:}  The systematic uncertainties reported are
estimated according to the following methods for the measurements of $v_3$ in both \pau and \dau collisions.

The effect of remaining background tracks due primarily to photon conversions and
weak decays is estimated by comparing the $v_3$ values when requiring a tighter
matching between the track projection and hits in PC3. We find that this
increases the $v_3$ by $<1\%$ and $7\%$ in \pau and \dau collisions, respectively,
independent of \pt.

The effect of double interaction event selection is estimated by comparing the $v_3$ values when
requiring a tighter cut on the rejection. This yields a change in the $v_3$ of
3\% and 2\% in \pau and \dau collisions respectively, independent of \pt.

Uncertainty in the event plane resolution comes from two sources. The first is
the statistical uncertainty inherent in the resolution calculation, which yields
a $\pm$13\% and $\pm$17\% uncertainty in \pau and \dau collisions, respectively.
Additionally, the resolution is calculated using central arm tracks over two
different \pt regions. This leads to an uncertainty of 7\% and 34\%
in \pau and \dau collisions, respectively.

We also include an uncertainty due to the choice of event plane detector. In \pau collisions, this is
determined by comparing the $v_3$ calculated using event planes determined by
the south BBC and FVTX. We find that the results are consistent within
uncertainties, as expected. In \dau collisions, $v_3$ is also calculated using an alternative
method utilizing two particle correlations. Based on a ratio of the $v_3$ values
calculated using the two particle correlation and event plane methods, we assign
a 16\% systematic uncertainty.

In $v_3$, nonflow decreases the amplitude of the measured signal~\cite{Adare:2015ctn}, and its contribution increases with increasing \pt.
To estimate the nonflow contribution we calculate a normalized correlation function between midrapidity tracks and BBC photomultiplier (PMT) tubes:
\begin{linenomath}
\begin{align}
S(\Delta \phi,\pt) &= \frac{d(Q_{\rm PMT}N^{\textrm{track}(p_T)-\textrm{PMT}}_{\textrm{same event}})}{d\Delta \phi}, \label{eq:S} \\
C(\Delta \phi,\pt) &= \frac{S(\Delta \phi,\pt)}{M(\Delta \phi,\pt)} \frac{\int_{0}^{2\pi}M(\Delta \phi,\pt)}{\int_{0}^{2\pi}S(\Delta \phi,\pt)}, \label{eq:C}
\end{align}
\end{linenomath}
where $Q_{\rm PMT}$ is the charge on the PMT in the pair and $N^{\rm track(\pt)-PMT}_{\textrm{same event}}$ is the number of track--PMT pairs from the same event. $M(\Delta 
\phi,\pt)$ is determined in the same way as $S(\Delta \phi,\pt)$
but with one particle in one event and another particle in a different event
(the so-called mixed event technique).  This normalization procedure accounts for
acceptance effects and produces a correlation function of order unity.
Next, we fit $C(\Delta \phi,\pt)$ with a Fourier expansion:
\begin{linenomath}
\begin{equation}
C(\Delta \phi ) = 1 + \sum 2 c_n(\pt)\cos(n\Delta \phi).
\label{eq:fphi}
\end{equation}
\end{linenomath}
We do this process for both systems in which we want to estimate the nonflow (\pau or
\dau) and for \pp at the same collision energy. We take the Fourier coefficients $c_n$ to
find the nonflow contribution to the $v_n$ values in a given system,
\begin{linenomath}
\begin{equation}
\textrm{nonflow ratio}=\frac{c_n^{\pp}(\pt) \frac{\langle Q^{\pp} \rangle}{\langle Q^{\rm system} \rangle}}{c_n^{\rm system}(\pt)}
\label{eq:cn}
\end{equation}
\end{linenomath}
where $\langle Q \rangle$ is the average BBC charge for the system. The ratio of average charges normalizes the $c_n$ by multiplicity. The assumption is that
$c_n^{\pp}$ is entirely due to nonflow such that the deviation of the nonflow ratio from one is taken as an estimate of the nonflow, and included as a \pt dependent systematic uncertainty.

\begingroup
\renewcommand{\arraystretch}{1.4} 
\begin{table*}[t] 
 \begin{tabular}{l c c c}
    \hline\hline
    Source & \pau & \dau & \hau \\
    \hline
    Track background & $\pm$4\% & $\pm$7\% & $\pm$5\% \\
    Event selection & $\pm$3\% & $\pm$2\% & $\pm$5\%\\
    $R(\psi_3)$ (sys.) & $\pm$7\% & $\pm$34\% & {\rm n/a} \\
    $R(\psi_3)$ (stat.) & $\pm$13\% & $\pm$17\% & {\rm n/a}\\
    $\psi_3$ determination & $<$1\% & $\pm$17\% & $\pm$15\%\\
    Detector alignment & $\pm$8\% & $\pm$5\% & $\pm$15\%  \\
    Nonflow (\pt dependent) & $+$21\%$\rightarrow$$+$114\% & $+$18\%$\rightarrow$$+$27\% & $+$4\%$\rightarrow$$+15$\%\\
    \hline
    \vspace{0.10mm}\\
    Combined & $_{-18\%}^{+27\%}\rightarrow_{-18\%}^{+115\%}$ & $^{+46\%}_{-42\%}\rightarrow^{+50\%}_{-42\%}$ & $^{+23\%}_{-22\%}\rightarrow^{+27\%}_{-22\%}$ \\
    \vspace{0.10mm}\\
    \hline\hline
  \end{tabular}
  \caption{Systematic uncertainties in the $v_3$ measurements as a function of \pt in 0-5\% central \pau, \dau, and \hau~\cite{Adare:2015ctn} collisions at $\sqsn=200$ GeV.}
  \label{tab:sys}
\end{table*}
\endgroup

A summary of the systematic uncertainties on $v_3$ in $p/d$$+$Au are given in Table~\ref{tab:sys} along with \hau uncertainties taken from Ref.~\cite{Adare:2015ctn}.


\textit{Comparison of theory to data:} The level of agreement between the 
different theoretical calculations and the data presented in this work is quantified
by performing a least squares fit incorporating a careful treatment of various 
types of systematic uncertainties, following Ref.~\cite{Adare:2008cg}.

The nonflow uncertainty is the dominant source of systematic uncertainty in
all six measurements. It is known to be point-to-point correlated as a function
of $p_T$, to contribute asymmetrically, and to be anti-correlated between $v_2$ 
and $v_3$. Namely, the nonflow can only reduce the measured $v_2$ 
while simultaneously only increasing the $v_3$. 

All remaining measurement uncertainties are assumed to be uncorrelated between
$v_2$ and $v_3$. The remaining uncertainties 
are assumed to contribute in the following ways:
\begin{enumerate}
	\item as point-to-point uncorrelated uncertainties
	\item as point-to-point anti-correlated uncertainties (e.g. a tilt in the $p_T$ dependence)
	\item as point-to-point correlated uncertainties
\end{enumerate}
The total systematic uncertainty (excluding the nonflow) is taken to contribute a fraction of its value 
to each of the above types. A conservative approach is taken, and these 
fractions are allowed to vary independently for each measurement within 
reasonable limits. 

The bands around the theoretical calculations shown in Fig.~\ref{fig:vn-models}
indicate some subset of theoretical uncertainties which differs between the
models. We
make the assumption that the dominant contribution is a point-to-point correlated
uncertainty which is additionally correlated between $v_2$ and $v_3$. Given their
small uncertainties, the inclusion of this treatment has little effect on the
results for either \sonic or MSTV. It has the largest effect with iEBE-VISHNU, 
however its inclusion does not affect the relative ordering of the agreement
discussed below.

We calculate a $p$-value from the least squares minimization in the standard way, 
where the number of degrees of freedom is simply the total number of data points, 
as there are no free parameters in the comparison. The total $p$-values, along
with the $p$-values for each collision system, are given in Table~\ref{tab:pval} 
for \sonic, iEBE-VISHNU, MSTV, and \ampt. The \ampt calculations are taken from 
Ref.~\cite{Koop:2015wea}, which calculate $v_2$ and $v_3$ relative to the initial 
participant nucleon plane, utilizing the so-called \textit{string melting} 
mechanism, and a parton interaction cross section of $\sigma=1.5$ mb. \sonic provides a very good description of the data, with 
a rather close to unity value of 0.96, which may indicate a modest overestimate of the statistical or systematic uncertainties. iEBE-VISHNU yields a worse $p$-value of 0.061. The larger $p$-value for \sonic compared to iEBE-VISHNU 
is driven by the nonflow uncertainty. The fact that \sonic tends to under-predict
the $v_2$ while over-predicting the $v_3$ is mitigated by the nonflow uncertainty, 
while iEBE-VISHNU's worse description of the $p_T$ dependence in \pau and \dau is 
not compensated for by the relatively small remaining uncertainty
Both MSTV and \ampt yield a very poor description of the data with $p$-values of 
$8.83\times10^{-17}$ and $1.71\times10^{-46}$ respectively.

\begingroup
\renewcommand{\arraystretch}{1.4} 
\begin{table}
	\caption{\label{tab:pval} Calculated $p$-values between model calculations
	and data.}
	\begin{tabular}{ccccc}
	\hline\hline
	 & \multicolumn{3}{c}{$p$-value} \\
	 & \sonic & iEBE-VISHNU & MSTV & \ampt \\
	\hline
	\pau     & 0.966 & 0.086 & $7.07\times10^{-17}$ & $1.88\times10^{-7}$ \\
	\dau     & 0.927 & 0.113 & 0.011                & $2.50\times10^{-21}$ \\
	\hau     & 0.465 & 0.385 & 0.007                & $6.67\times10^{-23}$ \\
	Combined & 0.960 & 0.061 & $8.83\times10^{-17}$ & $1.71\times10^{-46}$ \\
	\hline\hline
	\end{tabular}
\end{table}
\endgroup


\nolinenumbers
\bibliographystyle{naturemag}
\bibliography{main}



\end{document}